\documentclass[11pt, a4paper]{article}
\pdfoutput=1
\usepackage{hyperref}
\usepackage{amssymb,amsmath,amsfonts, mathtools, mathrsfs, doi, enumerate, cite}
\usepackage[utf8]{inputenc}
\usepackage[dvipsnames]{xcolor}
\usepackage{graphicx}
\usepackage{caption}
\usepackage{tensor}
\usepackage{subcaption}
\usepackage{enumerate}
\usepackage{dsfont}
\usepackage{comment}

\usepackage{verbatim}
\usepackage{amsfonts,euscript,amssymb,braket}
\usepackage{tikz}
\usetikzlibrary{arrows,decorations.markings,patterns}
\usepackage{slashed}

\def\clock{{\count0=\time
		\divide\count0 60
		\ifnum\count0<10 0\fi\the\count0
		\multiply\count0 -60 \advance\count0 \time
		:\ifnum\count0<10 0\fi \the\count0
}}
\newcommand{\timestamp}{{\small\vbox{\hbox{\tt\jobname.tex}
			\hbox{\the\day/\the\month/\the\year, \clock}}}}

\makeatletter
\let\old@startsection=\@startsection
\let\oldl@section=\l@section
\renewcommand{\@startsection}[6]{\old@startsection{#1}{#2}{#3}{#4}{#5}{#6\mathversion{bold}}}
\renewcommand{\l@section}[2]{\oldl@section{\mathversion{bold}#1}{#2}}
\makeatother

\numberwithin{equation}{section}

\begin{document}
	\renewcommand{\thefootnote}{\arabic{footnote}}

	\overfullrule=0pt
	\parskip=2pt
	\parindent=12pt
	\headheight=0in \headsep=0in \topmargin=0in \oddsidemargin=0in

	\vspace{ -3cm} \thispagestyle{empty} \vspace{-1cm}
	\begin{flushright} 
		\footnotesize
		\textcolor{red}{\phantom{print-report}}
	\end{flushright}

\begin{center}
	\vspace{1.2cm}
	{\Large\bf \mathversion{bold} Pseudo Complexity of Purification for free scalar field theories}	
	
	\vspace{0.5cm} {
		Aranya Bhattacharya$^{\, a,}$\footnote[1]{aranyab@iisc.ac.in},
		Arpan Bhattacharyya$^{\,b,}$\footnote[2]{abhattacharyya@iitgn.ac.in},	Sabyasachi Maulik$^{\,c,d,}$\footnote[3]{sabyasachi.maulik@saha.ac.in}
	}
	\vskip  0.5cm
	
	\small
	{\em
		$^{a}\,$Centre for High Energy Physics, Indian Institute of Science, C.V. Raman Avenue,\\ Bangalore 560012, India.
		\vskip 0.05cm
		$^{b}\,$Indian Institute of Technology, Gandhinagar, Gujarat 382355, India.
		\vskip 0.05cm
		$^{c}\,$Theory Division, Saha Institute of Nuclear Physics, 1/AF, Bidhannagar 700064,\\ West Bengal, India.
		\vskip 0.05cm
		$^{d}\,$Homi Bhabha National Institute, Training School Complex, Anushakti Nagar,\\ Mumbai 400094, India.
	}
	\normalsize	
\end{center}

\vspace{0.3cm}
\begin{abstract} 
We compute the \emph{pseudo complexity of purification} corresponding to the reduced transition matrices for free scalar field theories with an arbitrary dynamical exponent. We plot the behaviour of complexity with various parameters of the theory under study and compare it with the complexity of purification of the reduced density matrices of the two states $|\psi_1\rangle$ and $|\psi_2\rangle$ that constitute the transition matrix. We first find the transition matrix by reducing to a small number ($1$ and $2$) of degrees of freedom in lattice from a lattice system with many lattice points and then purify it by doubling the degrees of freedom ($2$ and $4$ respectively) for this reduced system. This is a primary step towards the natural extension to the idea of the complexity of purification for reduced density matrices relevant for the studies related to postselection.
\end{abstract}

\vspace{0.15 in}
\section{Introduction}\label{intro}
Quantum entanglement continues to grab eyeballs across multiple disciplines. Recently, a generalization of entanglement entropy: called \textit{pseudo entropy} was proposed in \cite{Nakata:2020luh, Mollabashi:2020yie}. There, instead of starting from a usual density matrix, the authors define a matrix using two pure quantum states $\ket{\psi_1}$ and $\ket{\psi_2}$
\begin{equation}
	\tau^{1\vert 2} = \frac{\ket{\psi_1}\bra{\psi_2}}{\langle \psi_2 \vert \psi_1 \rangle},
\end{equation}
which is dubbed the \textit{transition matrix}. This has a two-fold interest to begin with. The first is that in the case of post-selection in quantum experiments, the transition matrix becomes important once one specifies the initial state as $|\psi_1\rangle$ and the final state as $|\psi_2\rangle$, given they are not orthogonal to each other. In the post-selection experiment, the transition matrix then plays the role of the density matrix while computing the weak expectation value $\langle\mathcal{O}\rangle=\text{Tr}(\mathcal{O}\tau^{1|2})$ of an observable $\mathcal{O}$. The second motivation comes from the AdS/CFT correspondence \cite{Maldacena:1997re, Witten:1998qj}, where pseudo entropy is proposed to be the dual CFT quantity to the area of a minimal hypersurface in Euclidean AdS spacetime. In holography, this is constructed in the path integral technique by dividing a Euclidean timeslice into two regions corresponding to the two states. Dividing the entire system into $A$ and its complement $A^C$, the pseudo entropy of $A$ (relative to $A^C$) is defined as
\begin{equation}
	S\left(\tau^{1\vert 2}\right) = - \text{Tr}_{A} \left[\tau_A^{1\vert 2} \log\tau_A^{1\vert 2} \right],
\end{equation}
where $\tau_A^{1\vert 2} = \text{Tr}_{A^C}\left(\tau^{1\vert 2}\right)$ is called the \textit{reduced} transition matrix, in analogy with existing literature on entanglement entropy derived from reduced density matrix. Once the two states are taken to be the same, the transition matrix reduces to the usual density matrix and pseudo entropy reduces to usual entanglement entropy. 

Another quantum information (QI) theoretic  quantity that has been at the centre of interest for the last few years is quantum circuit complexity.  In recent times, circuit complexity has been explored in the context of quantum field theory \cite{Jefferson:2017sdb,Chapman:2017rqy, Hackl:2018ptj,Khan:2018rzm,Bhattacharyya:2018bbv,Bhattacharyya:2019kvj,me1,Magan:2018nmu,Caputa:2018kdj,Bhattacharyya:2019txx, Ali:2019zcj, Bhattacharyya:2020rpy,Bhattacharyya:2020kgu,Bhattacharyya:2020art, Erdmenger:2020sup,Flory:2020eot,Flory:2020dja,Erdmenger:2021wzc,Chagnet:2021uvi,Koch:2021tvp,Bhattacharyya:2022ren} \footnote{This list is by no means exhaustive. Interested readers are referred to the reviews \cite{Chapman:2021jbh, Bhattacharyya:2021cwf}, and references therein for more details.}. Complexity is usually defined as the number of elementary structural components needed to construct either an evolved state or operator from a simple initial state or operator. Here we will try to compute the circuit complexity of the reduced transition matrix. This reduced transition matrix is an analogue of a mixed state corresponding to a reduced density matrix, although it can be non-hermitian in general. The methods for computing circuit complexity we apply are those for the complexity of purification \cite{Ghodrati:2019hnn, Agn2019,Caceres:2019pgf,Camargo:2018eof,Camargo:2020yfv,Bhattacharyya:2020iic,Bhattacharyya:2021fii}. 
\par 
First, given a transition matrix, which is written in density matrix form, a state can be associated with this operator by writing the transition matrix in the vectorized form. Then we compute the circuit complexity for this state using Nielsen's method \cite{2005quant.ph..2070N, Nielsen_2006, 2007quant.ph..1004D}.  This vectorization means changing the bra part of the transition matrix to a ket which amounts to a doubling of the Hilbert space\cite{JAMIOLKOWSKI1972275,CHOI1975285,PhysRevA.87.022310}. Hence the corresponding transition state looks like
\begin{equation} \label{eq1}
\psi_{1|2}=\frac{|\psi_1\rangle|\psi_2\rangle}{\langle\psi_1|\psi_2\rangle},
\end{equation}
which, as mentioned in \cite{ Mollabashi:2020yie}, for a transition matrix, turns the corresponding state into a tensor product of the two states. Then the corresponding complexity, following the approach of \cite{Jefferson:2017sdb},  of the state dual to the transition matrix being 

\begin{equation} \label{eq2}
C(\tau^{1|2})=C(|\psi_1\rangle) \oplus C(|\psi_2\rangle).
\end{equation}

\noindent Note that, the complexity defined using canonical purification in (\ref{eq2}) is simply the direct sum of two complexities of two individual states $|\psi_{1,\, 2}\rangle$.  The direct sum is representative of the fact that the vectorized version is a tensor product. It is quite unlikely that this will have any nontrivial holographic interpretation, as typically holographic duals involve some sort of minimization.  Motivated by this, we want to define some notion of pseudo complexity of purification ($C(\tau_A^{1|2})$) in this paper extending the notion of complexity of purification \cite{Agn2019,Caceres:2019pgf,Camargo:2018eof,Camargo:2020yfv,Bhattacharyya:2020iic,Bhattacharyya:2021fii} for this case. We first purify the reduced transition matrix and compute the minimal complexity among all possible purifications \footnote{The state in (\ref{eq1}), can be obtained by using ``operator-state mapping," a particular example of purification, namely, ``canonical purification''.}. We do these computations for free scalar field theories and Lifshitz field theories in $\left(1+1\right)$ dimensions.

One can ask what pseudo complexity means physically. This is related to the idea of post-selection as suggested in \cite{Nakata:2020luh, Mollabashi:2020yie, Mollabashi:2021xsd} while defining pseudo entropy. The idea is that if, along some physical process, one starts with a state $|\psi_1\rangle$ and somehow the final state is post-selected to be $|\psi_2\rangle$, which is \emph{not} necessarily a result of a simple unitary evolution with a simple Hamiltonian, the process can be still approximated as a result of an operator $|\psi_2\rangle\langle\psi_1|$ operating on the initial state. In general, $|\psi_2\rangle$ might be a result of many unitary and non-unitary steps along the way \footnote{By non-unitary, we mean that there might be measurements made along an evolution due to which the state might collapse to a different state which results in an overall non-unitarity. It is worth noting that the process of purification by itself is a non-unitary one, which however is different from the non-unitarity caused by the post-selection.}. Hence, this operator $|\psi_2\rangle\langle\psi_1|$ might as very well be a non-hermitian operator. This operator is considered a transition matrix, similar to a density matrix in this treatment. The reason is that this operator does not have any nontrivial information of the system other than the outer product of the initial and the post-selected state, similar to a density matrix. Then the reduced transition matrix, defined after a reduction in degrees of freedom, can associate a notion of entropy similar to the entanglement entropy of a reduced density matrix. Similarly, the pseudo complexity is supposed to measure the complexity of this operator, whereas the pseudo complexity of purification is supposed to measure the complexity of forming such a reduced transition matrix once the two states $|\psi_{1,2}\rangle$ are given. Another point of view of looking at this problem is to relate it to the averaged number of maximally entangled pair of qubits to be distilled from the intermediate state once the final state is postselected \cite{Nakata:2020luh}. While pseudo entropy measures this number, pseudo complexity of purification can be thought to be measuring the amount of work (of course, in terms of resources or gates) needed for this distillation starting from the reduced transition matrix.

The rest of this article is organized as follows: in Sec.~(\ref{sec2}) we first review a path-integral calculation of Gaussian transition matrix following \cite{Mollabashi:2021xsd} and propose a definition of pseudo-complexity of purification. We then use our new definition and calculate this quantity in a simple setting of coupled harmonic oscillators on a lattice. The results of our numerical analyses are collected in Sec.~(\ref{sec3}). We conclude in Sec.~(\ref{sec4}) with discussions and outlook.  
\section{Pseudo Complexity of Purification: A Proposal}\label{sec2}

The concept of purification originates from quantum information theory (can be considered to be a close cousin of Schmidt decomposition as a process), and the process of purification refers to constructing a pure state from a mixed reduced density (/transition) matrix such that if one traces out the auxiliary degrees of freedom added while constructing the purified state, one gets back the original reduced density (/transition) matrix. While both the actual and purified state generates the same reduced density matrix, the purified state only has important information about the density matrix because the extra parameters added in the process of purification are random. The only constraints on them are that the purified state should follow the standard quantum mechanical postulates and properties of a pure state. As mentioned already, we will be considering random purification by all possible values of parameters for which the purified state can be consistently formed. However, we will choose a specific purification among them by minimizing the complexity functional \cite{Caceres:2019pgf}. This is motivated by the definition of complexity, where minimal resources play an important part. In terms of quantum gates, this means choosing the minimum number of gates so that resources needed are minimized. Now we describe our main setup.

\subsection{Setup}
We consider a system of scalar fields which is governed by the Hamiltonian
\begin{equation}
	H = \frac{1}{2}\int dx \left[\pi^2 + \left(\partial_x^z\right)^2 + m^{2 z}\phi^2 \right].
\end{equation}
This theory exhibits an anisotropic (Lifshitz) scaling symmetry $t \to \lambda^z t$, $x \to \lambda x$ in the $m \to 0$ limit; for $z=1$, this is an ordinary relativistic scalar field theory. For practial purposes, it is useful to consider the discretized Hamiltonian
\begin{equation}\label{Hamiltonian_discrete}
	H = \sum_{i=1}^{N} \left[ \frac{\pi_i^2}{2} + \frac{m^{2 z}}{2}\phi_n^2 + \frac{1}{2}\left(\sum_{k = 0}^{z} \left(-1\right)^{z+k} \begin{pmatrix}
		z \\ k
	\end{pmatrix} \phi_{i-1+k} \right)^2 \right],
\end{equation}
N being the total number of points in the lattice.\par

We consider a collection of coupled linear harmonic oscillators on a lattice of space points, labelled by capital Latin indices. Any two Gaussian states $\ket{\psi_\alpha}$ $\left(\alpha = 1, 2\right)$ of this system can be generically expressed in their position representation as
\begin{equation}
	\langle q_A\vert \psi_{\alpha}\rangle = N^{\left(\alpha\right)}\exp\left[-\frac{1}{2}q_A W_{AB}^{\left(\alpha\right)} q_B\right],
\end{equation}
where $q_A$ denotes displacement of the $A$-th oscillator, $W_{AB}^{\left(\alpha\right)}$ is a positive definite matrix and $N^{\alpha}$ is the normalization constant
\begin{equation*}
	N^{\left(\alpha\right)} = \left(\text{det}\left(\frac{W_{AB}^{\left(\alpha\right)}}{\pi}\right)\right)^{\frac{1}{4}}.
\end{equation*}
We divide the entire system into a part $A$ and its complement $A^C$, the lattice points within the sub-system $\Omega$ are labelled by lowercase Latin letters, while lowercase Greek indices label those outside it. We adopt the notation.
\begin{equation}
	W_{AB}^{\left(\alpha\right)} = 
	\begin{pmatrix}
		W_{ab}^{\left(\alpha\right)} & W_{a\beta}^{\left(\alpha\right)}\\
		W_{\alpha b}^{\left(\alpha\right)} & W_{\alpha\beta}^{\left(\alpha\right)}
	\end{pmatrix} = 
	\begin{pmatrix}
		A^{\left(\alpha\right)} & B^{\left(\alpha\right)}\\
		B^{\left(\alpha\right) T} & C^{\left(\alpha\right)}
	\end{pmatrix},
\end{equation}
to denote sub-matrices of $W_{AB}^{\left(\alpha\right)}$. The reduced transition matrix for the sub-system $\Omega$ in this representation is found by integrating over the rest
\begin{equation}
	\langle q_{a}\vert \text{Tr}_{A^C} \left(\ket{\psi_1}\bra{\psi_2} \right)\vert q_{b}\rangle = \int dq_{\alpha} \langle q_a, q_{\alpha}\vert \psi_1\rangle \langle \psi_2\vert q_b, q_{\alpha}\rangle.
\end{equation}
This is a Gaussian integral. Finally, the matrix elements of the reduced transition matrix can be expressed in a simple form \cite{ Mollabashi:2020yie}
\begin{equation}\label{red_tmat}
	\langle q^{\left(1\right)}\vert \tau_{A}^{1\vert 2}\vert q^{\left(2\right)} \rangle = \frac{N^{\prime}}{\sqrt{\text{det}\frac{\bar{C}}{\pi}}} \times \exp\left[-\frac{1}{2}
	\begin{pmatrix}
		q^{\left(1\right) T} & q^{\left(2\right) T}
	\end{pmatrix} M 
	\begin{pmatrix}
	q^{\left(1\right)}\\ q^{\left(2\right)}
	\end{pmatrix} \right]\,,
\end{equation}
where
\begin{equation}
	M = \begin{pmatrix}
		X^{\left(1\right)} & 2Y\\
		2Y^T & X^{\left(2\right)}
	\end{pmatrix}
\end{equation}
and
\begin{align}
	&X^{\left(\alpha\right)} = A^{\left(\alpha\right)} - \frac{1}{2}B^{\left(\alpha\right)} \bar{C}^{-1} B^{\left(\alpha\right) T}, \nonumber\\
	&Y = -\frac{1}{4}B^{\left(1\right)} \bar{C}^{-1} B^{\left(2\right) T},\nonumber\\\
	&\bar{C} = \frac{1}{2}\left(C^{\left(1 \right)} + C^{\left(2 \right)}\right), \nonumber\\\
	&N^{\prime} = \sqrt{\text{det}\frac{\bar{W}}{\pi}}, \nonumber\\\
	&\bar{W} = \frac{1}{2}\left(W^{\left(1\right)} + W^{\left(2\right)}\right)\,.
\end{align}

\subsection{Auxiliary parameters and purification: }

We use Nielsen's method for the  computation  of circuit complexity \cite{2005quant.ph..2070N, Nielsen_2006, 2007quant.ph..1004D, Jefferson:2017sdb} and choose the $F^1$ norm for the complexity functional \cite{Jefferson:2017sdb,Guo:2018kzl}. For a pure (or purified) state, this is written in terms of the (unentangled) reference state frequency and the frequency of the normalized version of the coupled oscillator system in the lattice pictures. We start from a various number of initial oscillators and reduce to transition matrices after tracing out all but one or two of the oscillators. Afterwards, we purify this reduced transition matrix $\tau_A^{1|2}$ by adding parameters corresponding to adding one or two more oscillators (similar to the doubling of degrees of freedom in qubits purifications). However, we do not choose canonical purification since it is always returns the original transition state  defined in (\ref{eq1}), the complexity of which is trivial, as explained before \footnote{As mentioned in (\ref{eq2}), for canonical purification the complexity of the purified state is simply the sum of the complexity of $|\psi_1\rangle$ and $|\psi_2\rangle$.}. For the reduced transition matrix, the parameters that come into the picture while matching the original reduced transition matrix with the reduced density matrix of the purified state take care of all possible purifications. In this optimal picture of purification, the number of unknown arbitrary parameters is much higher than the number of unknowns needed to purify $\rho_1^A$ or $\rho_2^A$. This is because the information of not one but two pure states goes into the transition matrix and hence the reduced transition matrix.

In order to define pseudo-complexity, we shall begin by introducing purification of the transition matrix $\tau_A^{1\vert 2}$. We consider a (fictional) auxiliary system $\tilde{A}$ whose Hilbert space is of the \textit{same dimension} as our original subsystem $A$ and consider two \textit{Gaussian states} in the enlarged Hilbert space $\mathcal{H}_A \otimes \mathcal{H}_{\tilde{A}}$
\begin{equation}
	\langle q_A, q_{\tilde{A}}\vert \Psi_{\alpha}\rangle = N_{A\tilde{A}} \exp\left[-\frac{1}{2}
	\begin{pmatrix} 
		q_A^T & q_{\tilde{A}}^T 
	\end{pmatrix}
 \begin{pmatrix}
 	J^{\left(\alpha\right)} & K^{\left(\alpha\right)}\\
 	K^{\left(\alpha\right) T} & L^{\left(\alpha\right)}
 \end{pmatrix}  
	\begin{pmatrix} 
		q_A\\ q_{\tilde{A}} 
	\end{pmatrix} \right], \hskip 0.5cm \left(\alpha = 1, 2\right)
\end{equation}
then proceeding exactly as above, we calculate a transition matrix in the enlarged Hilbert space. This particular choice of Gaussian purification is primarily motivated by the study done in \cite{Bhattacharyya:2018sbw} in the context of entanglement of purification. This simplifies the numerical analysis as well as the computation of circuit complexity. We consider that this is a purification of our original reduced transition matrix \eqref{red_tmat} for the subsystem $A$; in this case, the following constraints must be obeyed.
\begin{align}
	&J^{\left(\alpha\right)} - \frac{1}{2}K^{\left(\alpha\right)}\bar{L}^{-1}K^{\left(\alpha\right) T} = A^{\left(\alpha\right)} - \frac{1}{2}B^{\left(\alpha\right)} \bar{C}^{-1} B^{\left(\alpha\right) T},\\
	&K^{\left(1\right)}\bar{L}^{-1}K^{\left(2\right) T} = B^{\left(1\right)} \bar{C}^{-1} B^{\left(2\right) T},
\end{align}
where $\bar{L} = \frac{1}{2}\left(L^{\left(1\right)} + L^{\left(2\right)}\right)$. In our numerical analysis, we adopt the following conditions to satisfy these constraints
\begin{subequations} \label{JKL_eqs}
\begin{align}
	&J^{\left(\alpha\right)} = A^{\left(\alpha\right)}, \label{JKL_eqs1} \\
	&\bar{L} = \left(K^{\left(2\right) -1} B^{\left(2\right)} \bar{C}^{-1} \left(K^{\left(2\right)} B^{\left(2\right)} \right)^T \right)^{-1}, \label{JKL_eqs2} \\
	&K^{\left(2\right)} = \left(\bar{L} K^{\left(1\right) -1} B^{\left(1 \right)} \bar{C}^{-1} B^{\left(2 \right) T} \right)^T\,. \label{JKL_eqs3}
\end{align}
\end{subequations}
This particular choice leaves only $K^{\left(1 \right)}$ and $L^{\left(1 \right)}$ undetermined.

For the particular model of our choice, the matrices $W_{AB}^{\left(\alpha \right)}$ are given by
\begin{equation}
	W_{AB}^{\left(\alpha \right)} = \frac{1}{N} \sum _{C=1}^{N} \sqrt{m_{\alpha}^{2 z_{\alpha}} + \left(2 \left(1 - \cos\frac{2\pi n}{N} \right) \right)^{z_{\alpha}}} \exp\left(\frac{2\pi i C \left(A-B \right)}{N} \right), \hskip 0.5 cm \left(\alpha = 1, 2 \right),
\end{equation}
where $N$ is the total number of lattice points. $m_{\alpha}$ and $z_{\alpha}$ are the mass and dynamical exponent of the respective theory. The matching of $J,\, K,\, L^{(\alpha)}$ matrices with their corresponding counterparts introduce a few arbitrary parameters in the purified state. These parameters denote the infinitely many possible Gaussian purifications of the reduced transition matrix. Afterwards, we define the complexity of purification as the complexity of the specific purification for which the complexity functional is minimized in terms of all the parameters of the purified version of the reduced transition matrix. This involves  purification parameters for both $|\psi_{1,\, 2}\rangle$ states, constituting the transition matrix. Finally, we compare the complexity of purification of the reduced transition matrix $C(\tau_A^{(1|2)})$ with the complexity of purifications for the individual reduced density matrices ($\rho_A^{1}$ and $\rho_A^{2}$) derived from the initial and final states $|\psi_1\rangle$ and $|\psi_2\rangle$ respectively.\par

We propose that the pseudo complexity of purification  is given by
\begin{equation} \label{pseudo_CoP_defn}
	\mathcal{C}_P = \underset{\tilde{A}}{\mathrm{min}} \frac{1}{2}\sqrt{\sum_{j=1}^{N}\sum_{\alpha=1}^{2}\log\left(\frac{\Omega^{\left(\alpha\right)}_j}{\omega_2}\right)},
\end{equation}
where $\Omega^{\left(\alpha\right)}_j$ happens to denote the $j$-th eigenvalue of the matrix $\begin{pmatrix}
	J^{\left(\alpha\right)} & K^{\left(\alpha\right)}\\
	K^{\left(\alpha\right) T} & L^{\left(\alpha\right)}
\end{pmatrix}$, the functional is minimized over the parameters associated with the auxiliary system $\tilde{A}$. This means choosing a particular purification associated with minimizing parameter values.

Let us point out a few attributes of our method of computation. Most of our numerical analyses are performed for the simplest case of a reduced system made of only one oscillator; where the above functional is to be minimized over two unknown parameters. In general, when the reduced system consists of $n$ no. of linear harmonic oscillators; the $J$, $K$, and $L$ matrices are of dimension $n \times n$. The choice \eqref{JKL_eqs} always leaves two of them undetermined, and therefore we end up with a total of $2n^2$ unknown parameters. On the other hand, to calculate the more well-known complexity of purification from a reduced density matrix, the no. of unknown parameter values over which one is required to perform an minimization scales as $n^2$ with the reduced system size. It is easy to conclude that the calculation of the new functional \eqref{pseudo_CoP_defn} is at least twice as hard since the transition matrices have information of both the initial as well as the final state. 

There, however, exists an way to purify the system mode-by-mode \cite{Caceres:2019pgf}. That way it might be possible to reduce the no. of unknown parameters to $2n$. Even then, the problem is harder (twice) than computing complexity of purification from density matrices, where the associated no. scales as $n$.

It may also be of relevance to note that any purification we consider takes into account only Gaussian states and the auxiliary system always has the same dimension as the original one, motivated by the usual doubling of Hilbert space typically done in purification using Schmidt decomposition. These are restrictions we invoke in order to keep the numerical job simple. There may exist ways to relax the conditions.

\section{Results and discussion} \label{sec3}

\subsection{On the  pseudo complexity of purification: }

In this subsection, we report our numerical results after the minimization with varying different variables of the theory. In the primary Hamiltonian \eqref{Hamiltonian_discrete}, we chose a general version which applies to both the ordinary free scalar QFTs $(z_1=z_2=1)$ as well as Lifshitz scalar field theories $(z_1 \neq 1 \neq z_2)$. The variables considered are the dynamical exponents $z_{1,2}$, the masses $m_{1,2}$, the number of oscillators before tracing out and the number of oscillators of the reduced system and the frequency of the unentangled reference state.

\begin{enumerate}

   \item \textbf{Varying $z_2$ and $m_2$:} From the plots of pseudo complexity of purification against reference frequency in Figs.~(\ref{diff_z2} )and (\ref{diff_m2}), we notice that as the reference frequency is increased, the pseudo complexity of purification decreases primarily and then shows a polynomial growth and saturation. When the scaling factor $z_2$ is increased in the same plots, the pseudo complexity of purification for  higher values of reference frequency decreases for higher $z_2$. However, the behaviour is  opposite for small reference frequency values. Hence, there is a crossover in complexity after which the small to large behaviour changes. Similar change is observed from the Fig.~(\ref{diff_m2}) where the mass parameter $m_2$ is varied. We find that at large frequencies, the pseudo  complexity of purification decreases as the mass is increased, whereas, at very small frequencies, the behaviour is again opposite.
   \begin{figure}[t]
   	\begin{subfigure}{0.498\textwidth}
   		\centering
   		\includegraphics[width=\textwidth]{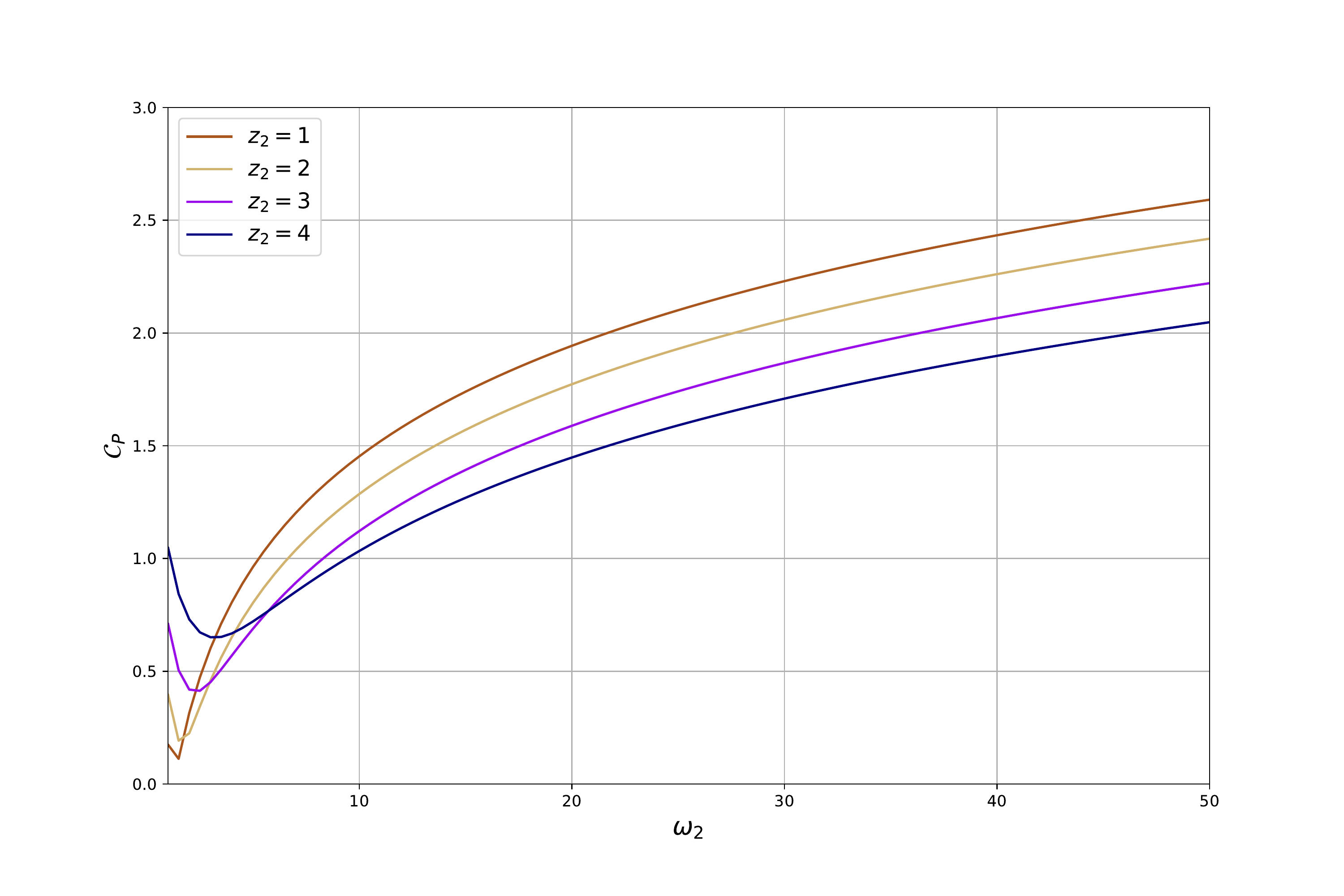}
   		\caption{}
   		\label{diff_z2}
   	\end{subfigure}
   	\hfill
   	\begin{subfigure}{0.498\textwidth}
   		\centering
   		\includegraphics[width=\textwidth]{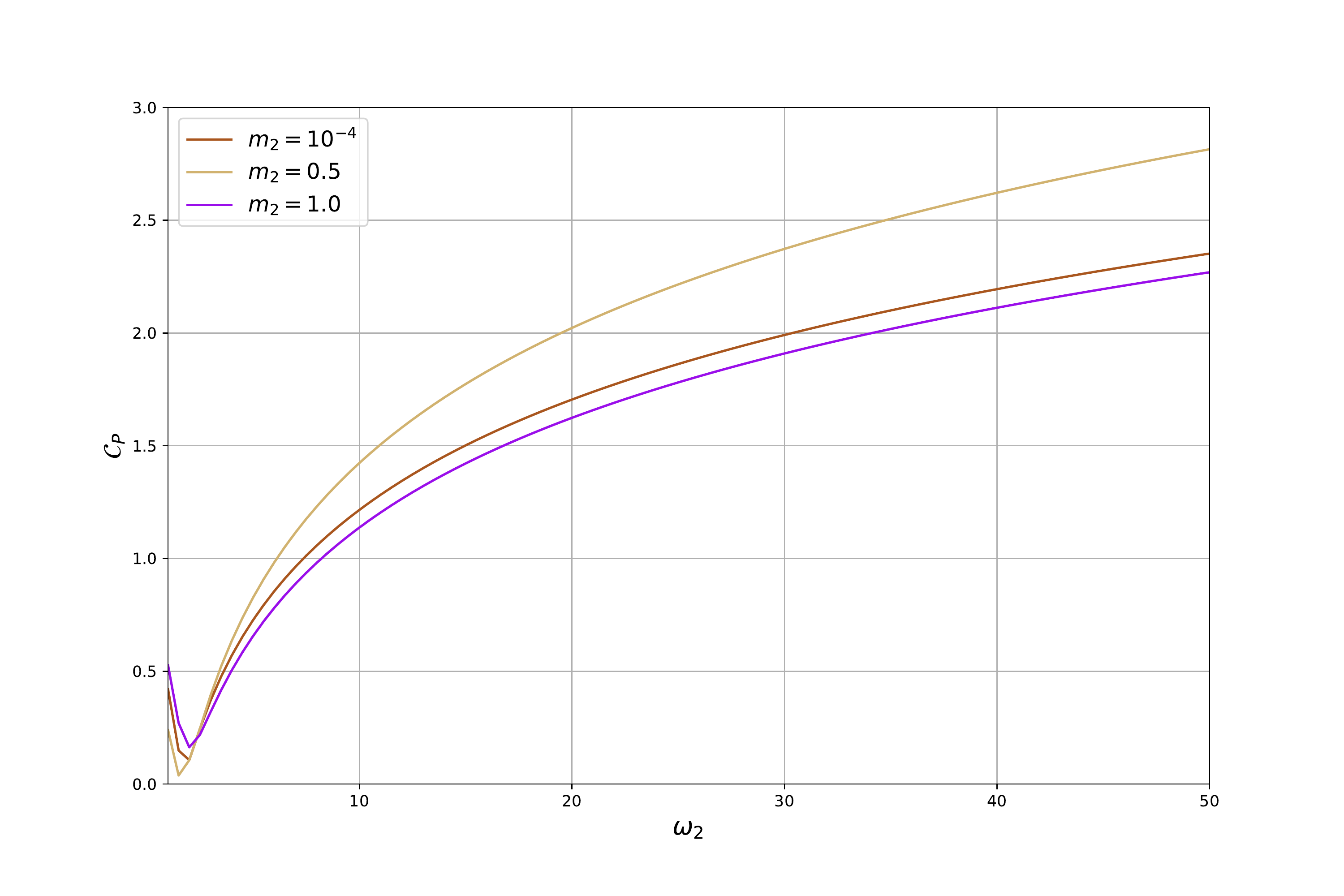}
   		\caption{}
   		\label{diff_m2}
   	\end{subfigure}
   	\caption{Behaviour of $\mathcal{C}_P$ with increasing $\omega_2$ for (a) $m_1 = m_2 = 0.5, z_1 = 1$ and different $z_2$ and (b) $m_1 = 1.0, z_1 = 1, z_2 = 2$ and different $m_2$.}
   \end{figure}

   \item \textbf{Pseudo complexity of purification vs $z_2$ plots:} From Fig.~(\ref{dependence_z2}) we observe that as $z_2$ is increased, the pseudo complexity of purification primarily goes down and then grows linearly. By looking at multiple plots for various values of $z_1$, we find that with increasing $z_1$ as $z_2$ is varied, the pseudo complexity of purification decreases. In this case, there is no change of behaviour or crossover between different plots, which we saw earlier. 
   
   \item \textbf{Pseudo complexity of purification vs $m_2$ plots:} As shown in Fig.~(\ref{dependence_m2}), we find that one of the mass parameters is varied from zero to one, and the pseudo  complexity of purification does not change much. Although the plots are not completely parallel to the x-axis, their variance is relatively much smaller than what we have observed for other parameters. On the other hand, if we vary one of the scaling parameters $z_2$, we find that with increasing values of $z_2$, pseudo complexity of purification increases.
   \begin{figure}[t]
   	\begin{subfigure}{0.498\textwidth}
   		\centering
   		\includegraphics[width=\textwidth]{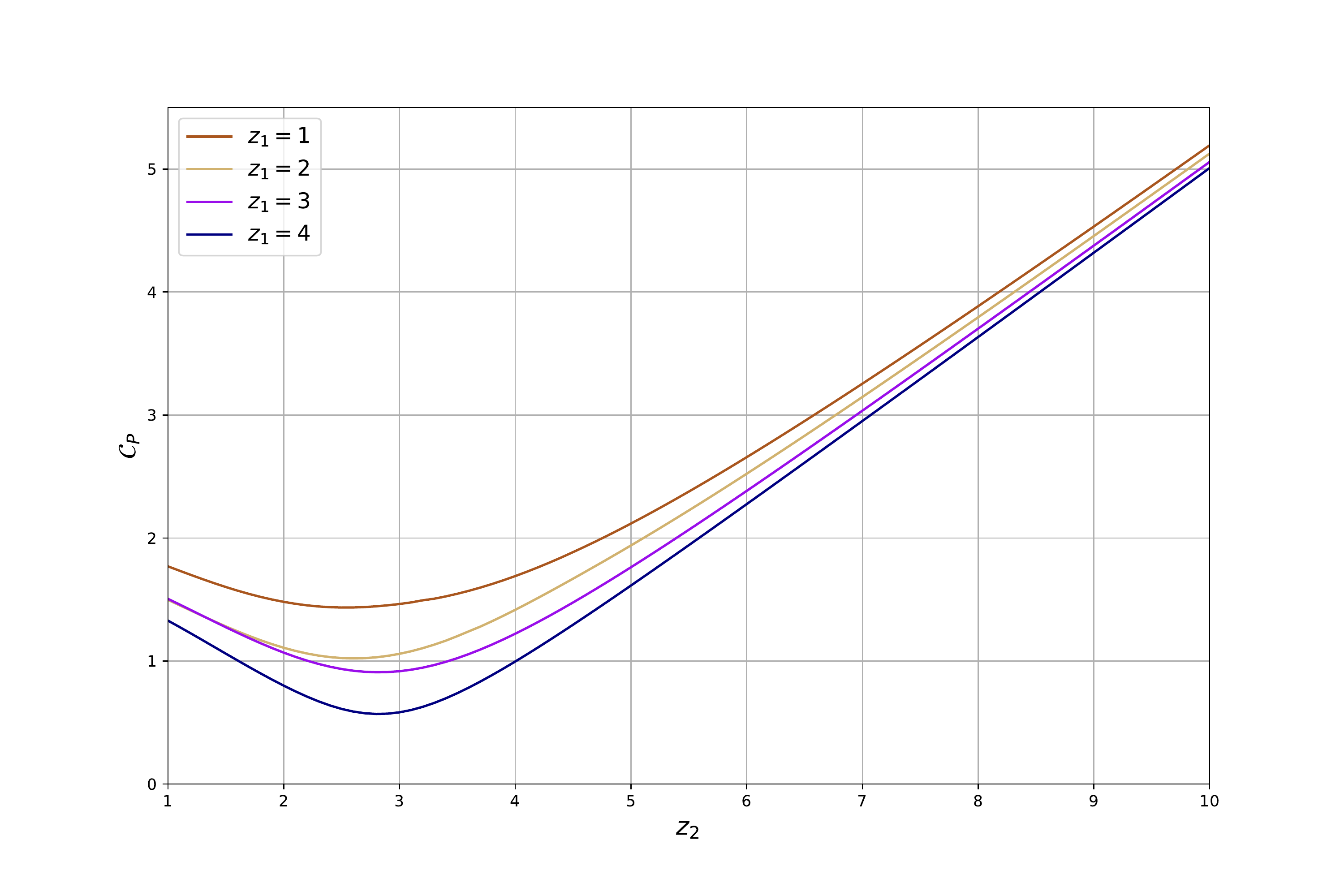}
   		\caption{}
   		\label{dependence_z2}
   	\end{subfigure}
   	\hfill
   	\begin{subfigure}{0.498\textwidth}
   		\centering
   		\includegraphics[width=\textwidth]{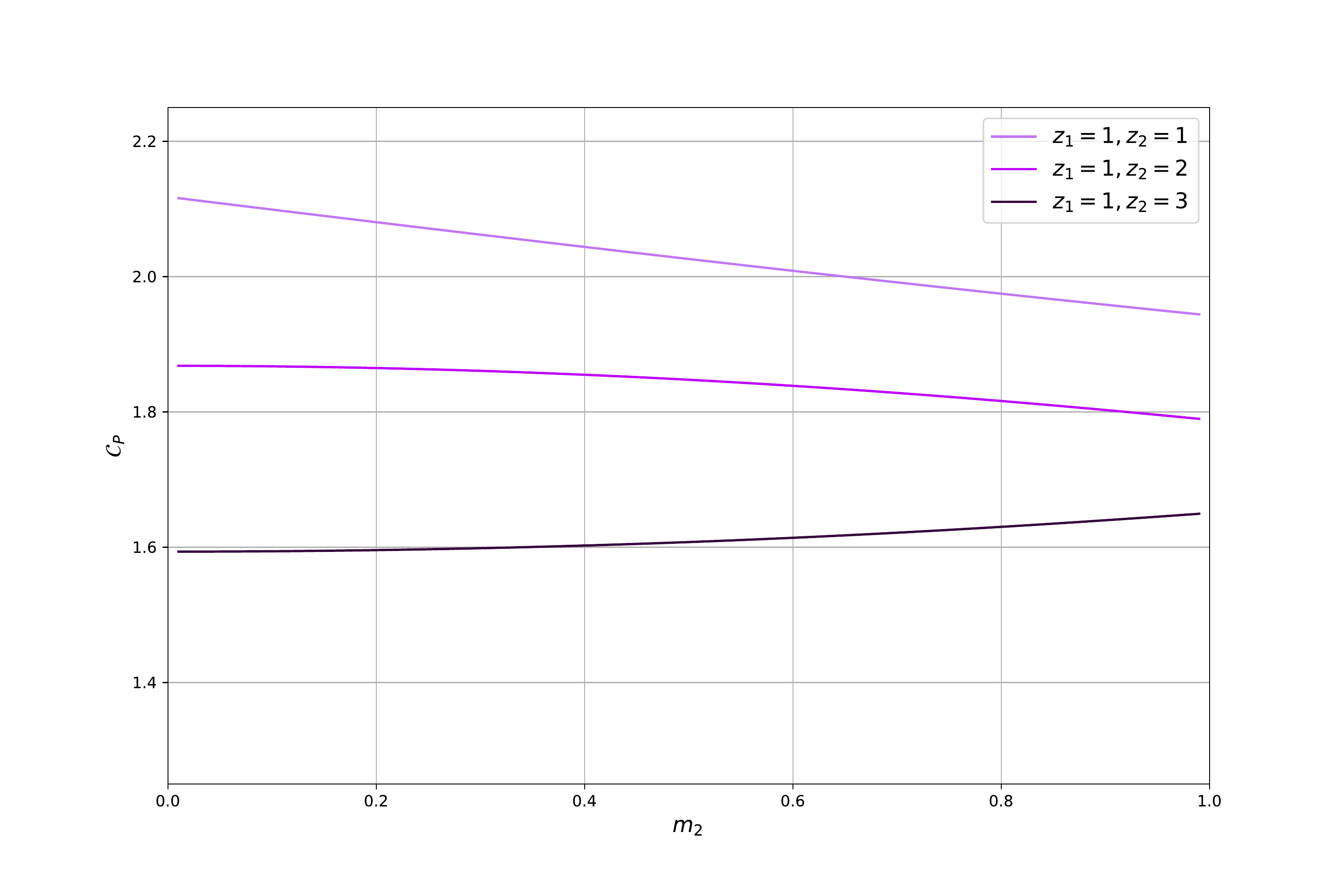}
   		\caption{}
   		\label{dependence_m2}
   	\end{subfigure}
   	\caption{Behaviour of $\mathcal{C}_P$ with increasing (a) $z_2$ and (b) $m_2$ for $\omega_2 = 25$.}
   \end{figure}

   \item \textbf{Varying number of oscillators before tracing out:} All the above plots are given for the case where the total number of oscillators associated with the states $\psi_{1,2}$ before tracing out degrees of freedom is $2$. So we trace out one of the two oscillators in each of the states. This begs the question of how the results might possibly change in case the number of oscillators are increased, which is more relevant when thinking about quantum field theories, which can be written as the lattice of infinitely many coupled harmonic oscillators. We, however, find that the pseudo complexity of purification does not vary much at all when the number of oscillators of the actual system is varied. This can be seen from Fig.~(\ref{diff_N}).
   \begin{figure}[t]
   	\centering
   	\includegraphics[scale=0.3]{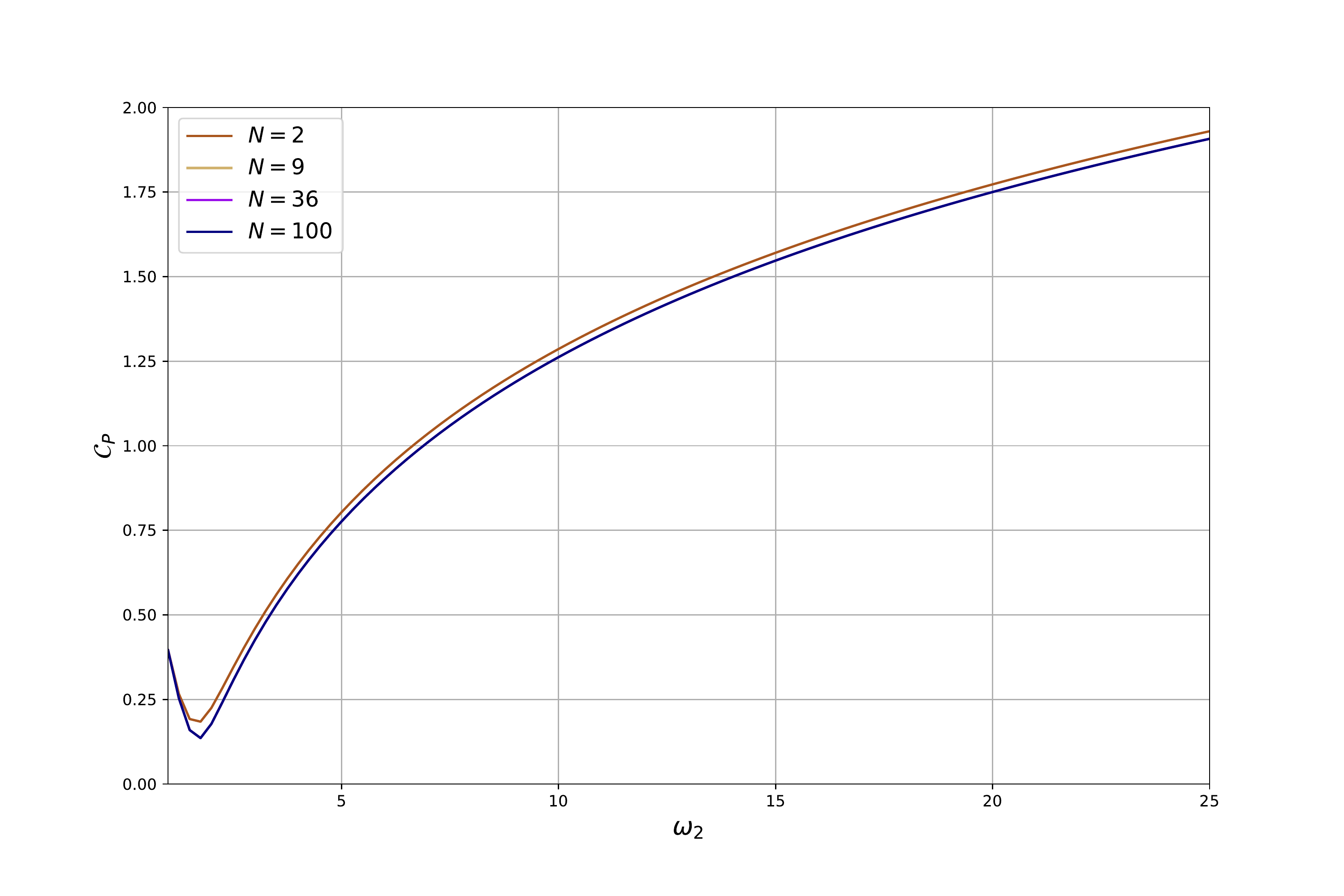}
   	\caption{Comparison of $\mathcal{C}_P$ for different total number of lattice sites $\left(N\right)$. In all graphs $m_1 = m_2 = 0.5$ and $z_1 = 1$, $z_2 = 2$.}
   	\label{diff_N}
   \end{figure}

   \item \textbf{Varying number of oscillators in the reduced system:} All of our results so far are for the case when the number of oscillators in the reduced transition matrix is one. The reason is two-fold: first and foremost, as we have mentioned before, the number of unknown parameters for pseudo complexity of purification in terms of which the complexity functional is minimized is much more than the case of usual complexity of purification. This makes the problem much more challenging numerically as the number of parameters grow very quickly as one increases the number of oscillators in the reduced system. Secondly, from a physical point of view, with the motivation of relating the entanglement entropy to some thermal notion of entropy, one typically chooses the number of oscillators in the reduced system to be small. Although there is no such clear motivation for complexity, we stick to  this particular choice.
   
	The qualitative behaviour of the pseudo complexity of purification with the varying reference frequency remains similar when the subsystem size is increased to include two oscillators as shown in Fig.~(\ref{CoP2plus2}). Hence, we expect all of our results to be universal and hold for larger subsystem size.
   \begin{figure}[t]
   	\centering
   	\includegraphics[scale=0.40]{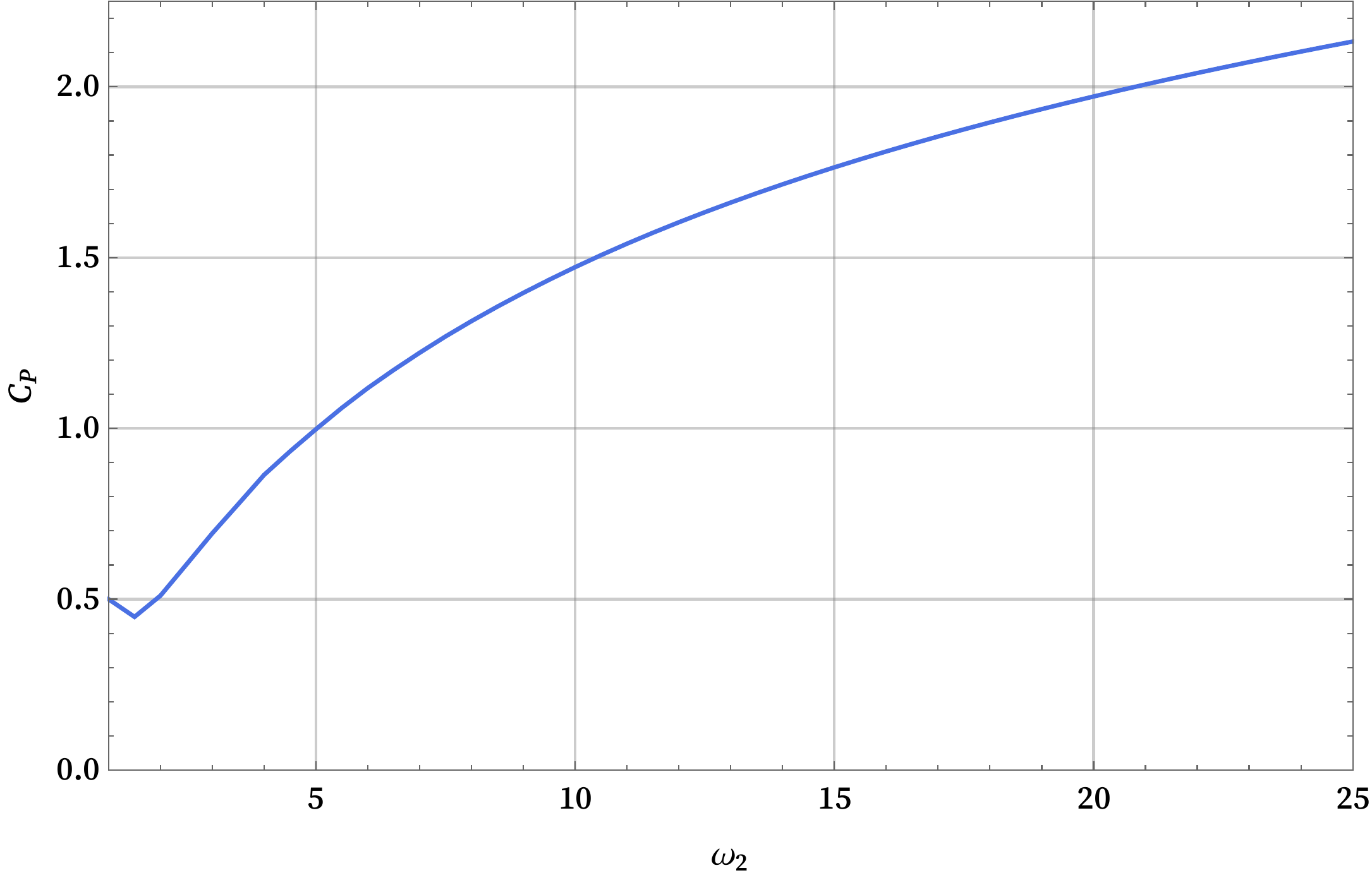}
   	\caption{Pseudo complexity of purification $\mathcal{C}_P$ when the reduced system contains two LHOs.}
   	\label{CoP2plus2}
   \end{figure}
   
\end{enumerate}

\subsection{On the difference $\Delta C$: Mutual pseudo complexity of purification} \label{mutualcop}

The reduced transition matrix carries complicated mixed information of not one but two pure states $\psi_1$ and $\psi_2$. Hence, while studying the complexity of purification for the reduced transition matrix, it is natural to ask how different it is from the complexities of individual reduced density matrices of those two states. Inspired by the definition of mutual complexity $\left(\triangle C\left(\rho_{AB}\right)\right)$ \cite{Alishahiha:2018lfv, Caceres:2018blh, Caceres:2019pgf}, which is defined as the difference of complexity between the full state $\rho_{AB}$ and the sum of complexities of the two reduced density matrices $\rho_A$ and $\rho_B$, we define the difference as the \emph{mutual pseudo complexity of purification}. In the following, we report some properties of the mutual pseudo complexity of purification based on our  numerical analysis.
\begin{enumerate}
   \item \textbf{Subadditivity:} From figure \ref{comparison_CoP}, it is easy to observe that 
   \begin{equation}
   \triangle C\left(\tau_A^{1|2}\right) = C\left(\tau_A^{1|2}\right)-\frac{C\left(\rho_A^1\right)+C\left(\rho_A^2\right)}{2} \leq 0.
   \end{equation}
   Therefore, we find that the pseudo complexity of purification is always sub-additive to the sum of the individual complexities of purification of the two relevant states. Our numerical results suggest that the term $\triangle C\left(\tau_A^{1|2}\right)$ is always negative or zero. The exact equality is expected for the case when the two states become the same and hence $\rho_A^1=\rho_A^2$.
   \begin{figure}[t]
   	\begin{subfigure}{0.498\textwidth}
   		\centering
   		\includegraphics[width=\textwidth]{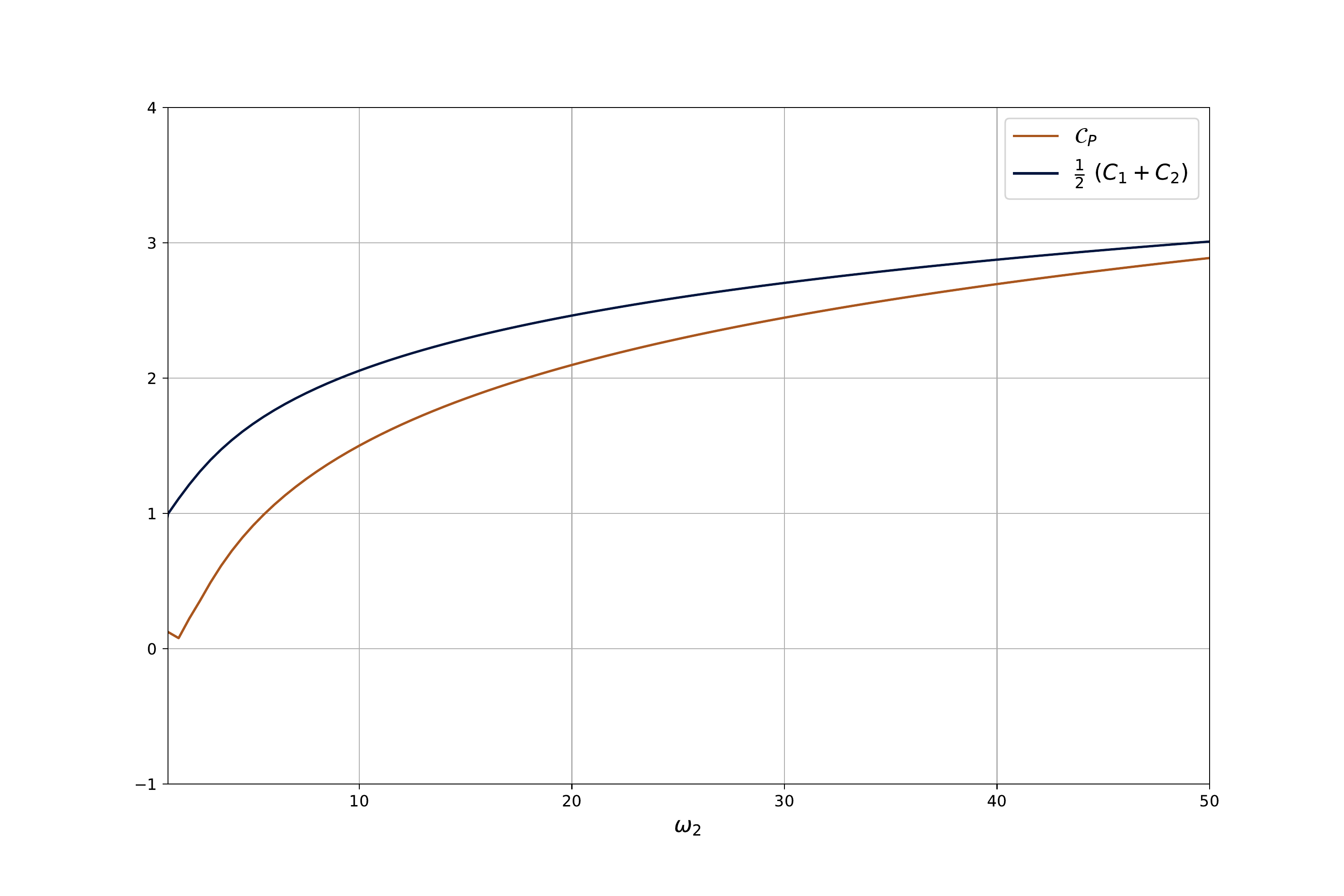}
   		\caption{}
   	\end{subfigure}
   	\hfill
   	\begin{subfigure}{0.498\textwidth}
   		\centering
   		\includegraphics[width=\textwidth]{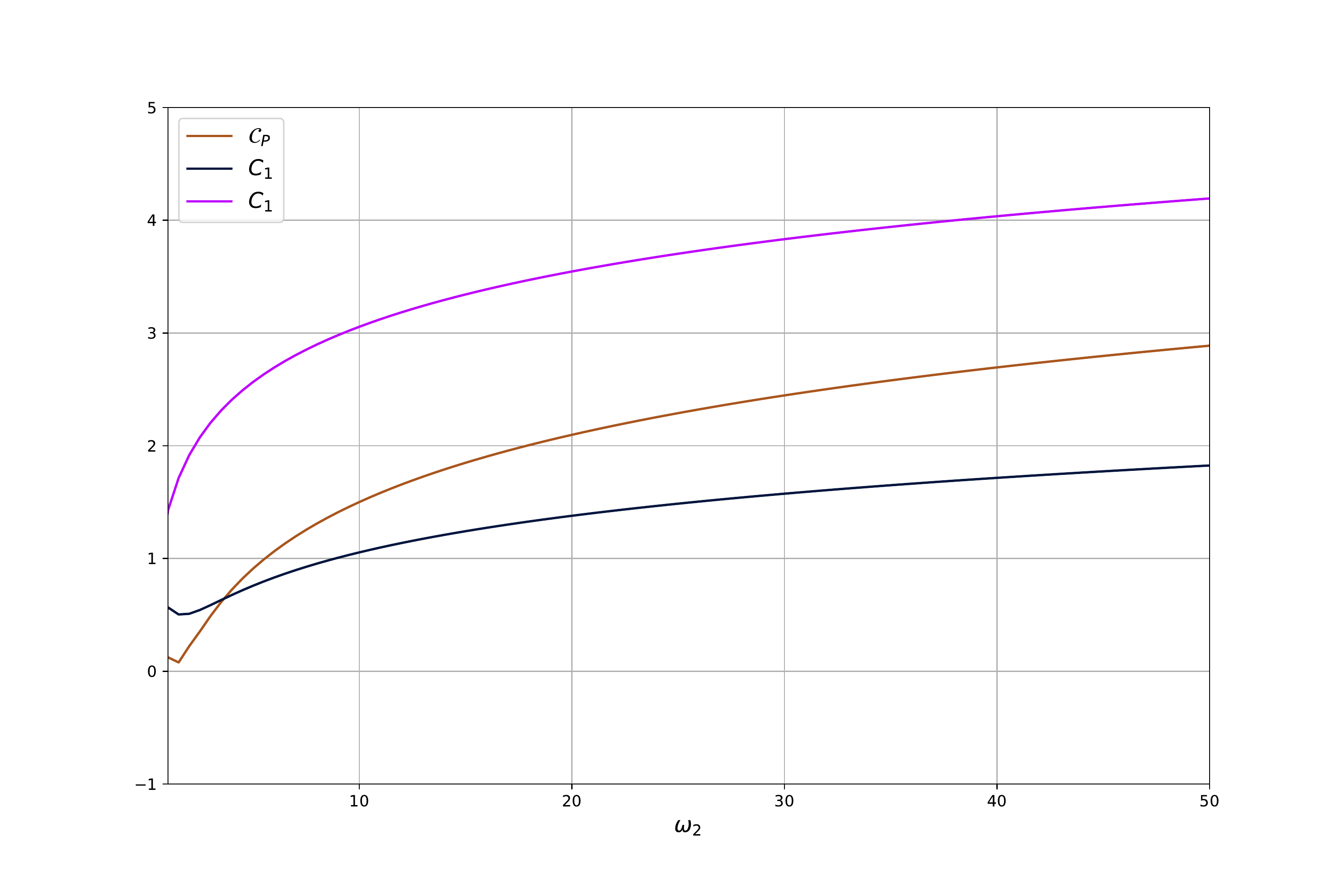}
   		\caption{}
   	\end{subfigure}
   	\caption{Comparison of $\mathcal{C}_P$ with individual complexities of purification of the two relevant states $\psi_1$ and $\psi_2$, we let $z_1 = 1$, $z_2 = 2$ and $m_1 = m_2 = 0.5$.}
   	\label{comparison_CoP}
   \end{figure}

   \item \textbf{Near saturation behaviour:} It is observed that both the complexity of purification and the pseudo complexity of purification tend to saturate at a very large reference frequency. By looking at the saturation values of the corresponding plots in Fig.~(\ref{comparison_CoP}), we can make the following comment on the saturation of pseudo complexity of purification in terms of complexities of purification of the individual reduced density matrices,
   \begin{equation}
   C(\tau_A^{1|2})\sim \frac{C(\rho_A^1)+C(\rho_A^2)}{2}   
   \end{equation}
   This means that the quantity $\Delta C(\tau_A^{1|2})$ approaches zero at large reference frequencies. This indicates that the dependence of $C(\tau_A^{1|2})$ on the reference frequency  $\omega$ is such that for large enough values of $\omega$, the complexity does not distinguish between a transition matrix and the density matrix.  
   
   \item \textbf{Varying masses:} We also study the behaviour of $\frac{C(\rho_A^1)+C(\rho_A^2)}{2} -C(\tau_A^{1|2})$ ($= \triangle C(\tau_A^{1|2})$) with change in one of the masses $m_2$ corresponding to the state $\psi_2$ while keeping the other parameters ($m_1,\, z_1, \, z_2$) fixed. We look at the behaviour both in a small mass difference as well as a large mass difference regime. In both cases, we find that the difference decreases as one increases the difference between mass parameters, see Fig.~(\ref{comparison_mass_CoP}). In the small mass difference regime, the slope of the plot is linear. This means $\left|\triangle C(\tau_A^{1|2}) \right| \propto -a \left(m_2-m_1 \right)$ in this range. On the other hand in the large mass difference regime, the decay of the slope is close to exponential inspiring us to write $\left|\triangle C(\tau_A^{1|2})\right| \propto e^{-a \left(m_2-m_1 \right)}$, with lower saturation close to zero as $\left(m_2-m_1 \right)\sim 1$.
   \begin{figure}[t]
   	\begin{subfigure}{0.498\textwidth}
   		\centering
   		\includegraphics[width=\textwidth]{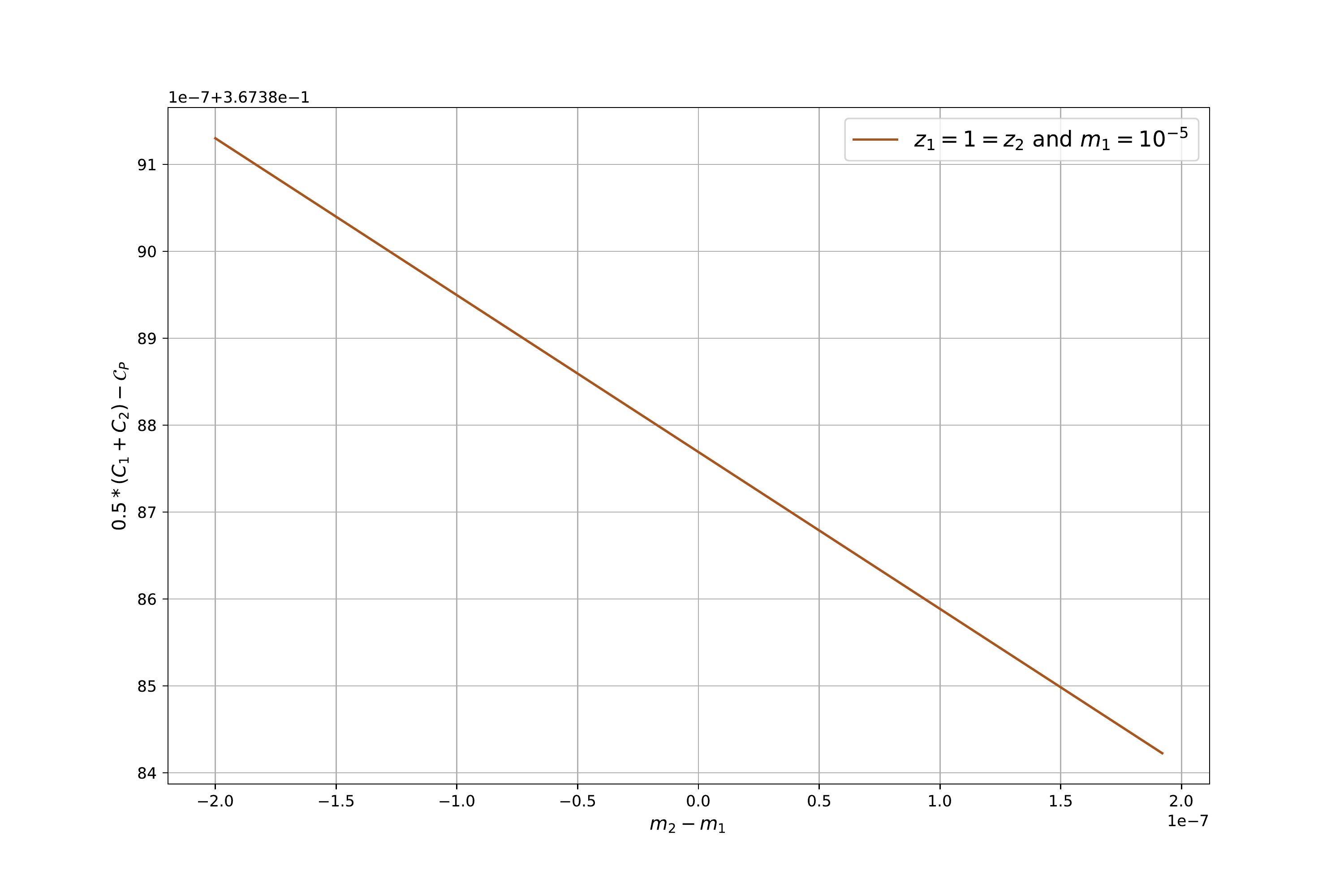}
   		\caption{}
   	\end{subfigure}
   	\hfill
   	\begin{subfigure}{0.498\textwidth}
   		\centering
   		\includegraphics[width=\textwidth]{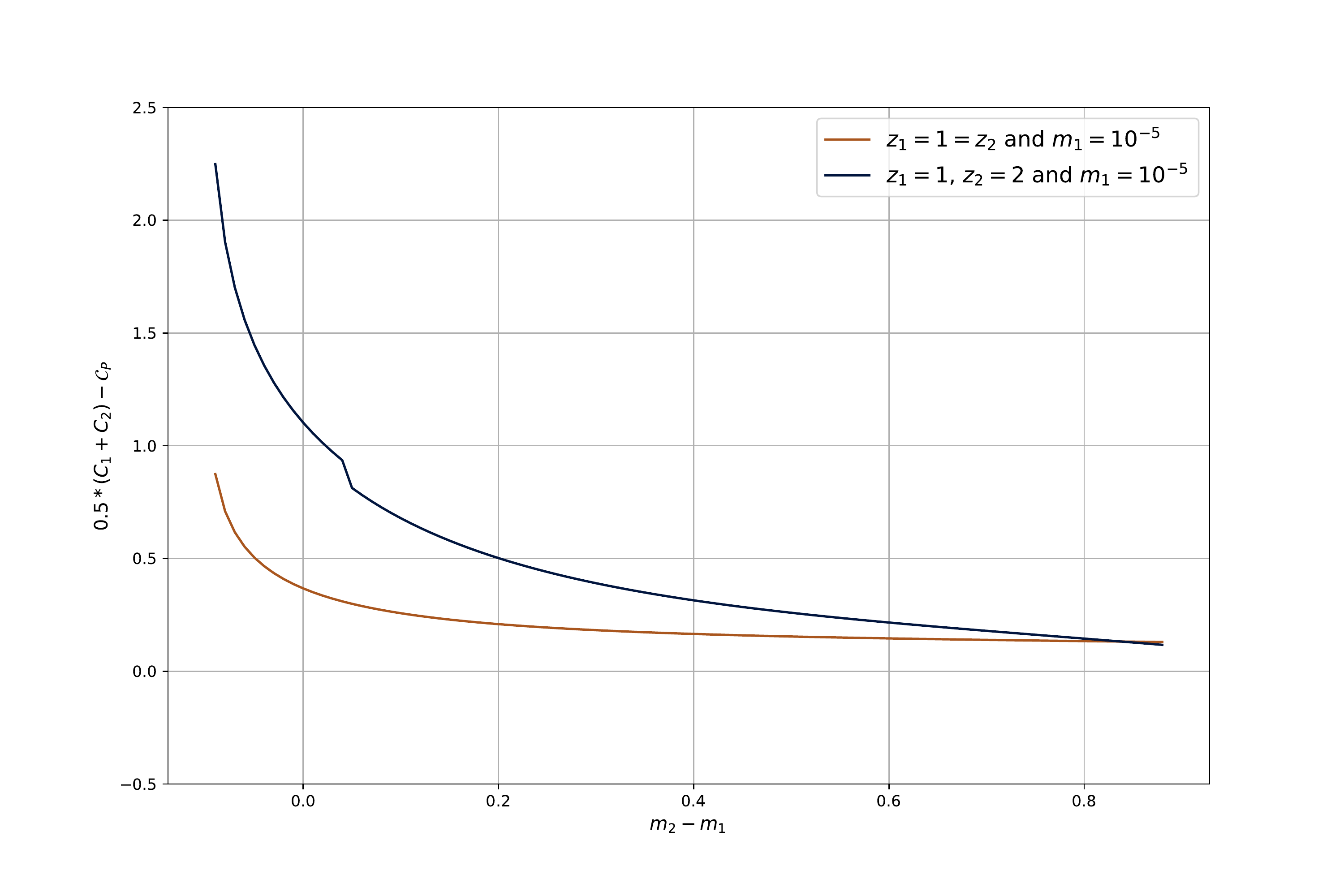}
   		\caption{}
   	\end{subfigure}
   	\caption{Behaviour of $\frac{1}{2}\left(C_1 + C_2\right) - \mathcal{C}_P$ (a) small mass difference and (b) large mass difference. In all results, $\omega_2 = 50$.}
   	\label{comparison_mass_CoP}
   \end{figure}

   \item \textbf{Varying $z_2$:} In the same figure, we see that as the difference between $z_2-z_1$ is increased, the absolute value of $\triangle C(\tau_A^{1|2})$ increases. However, the lower saturation $\left|\triangle C\left(\tau_A^{1|2}\right) \right| \sim 0$, near $(m_2-m_1)\sim 1$, remains unchanged. Hence, the numerics definitely suggests that the saturation is universal and independent of $z_2-z_1$.
    
   \item \textbf{ $\left|\triangle C\right| \, \text{vs}\, z_2$ plot:} As shown in Fig.~(\ref{comparison4_z2depdn_CoP}), if we keep $z_1,\, m_1, \, \text{and} \, m_2$ fixed while varying $z_2$, $|\triangle C| $ increases linearly. 
   \begin{equation}
   |\triangle C| \propto \alpha z_1+\beta  z_2,
   \end{equation}   
   the above expression is written in this way because our treatment is symmetric with respect to $z_{1,2}$. In the same plot, we also notice that once $z_1$ is chosen to be $2$ instead of $1$, $|\triangle C|$ increases, but the linear behaviour remains unchanged.
   \begin{figure}[t]
   	\centering
   	\includegraphics[scale=0.35]{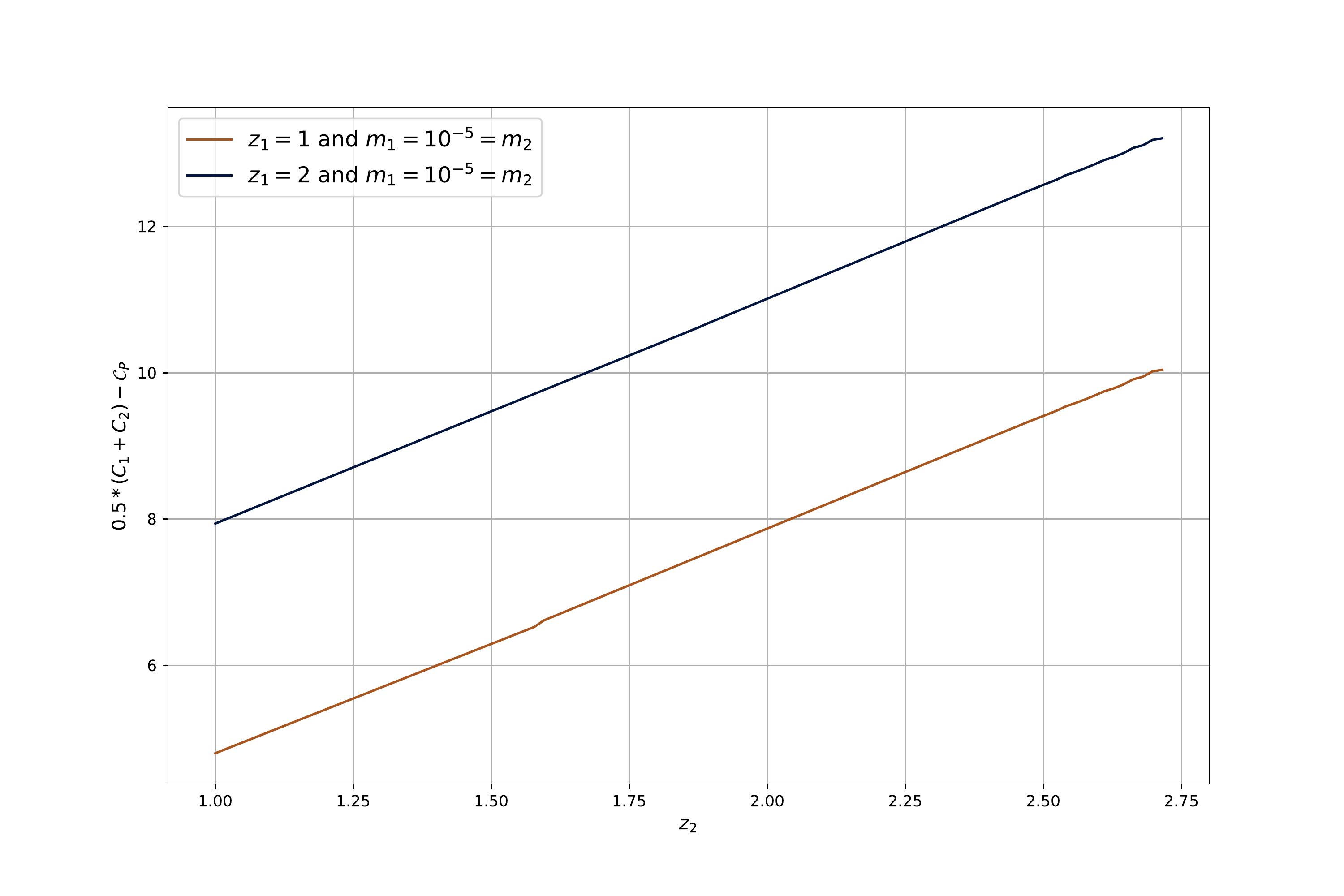}
   	\caption{Behaviour of $\frac{1}{2}\left(C_1 + C_2\right) - \mathcal{C}_P$ with $z_2$ in the small mass regime.}
   	\label{comparison4_z2depdn_CoP}
   \end{figure}
   \end{enumerate}

\section{Conclusion}\label{sec4}
We have studied the pseudo complexity of purification of the reduced transition matrix \eqref{pseudo_CoP_defn} numerically for free scalar and Lifshitz scalar field theories. We also defined the mutual pseudo complexity of purification by comparing the pseudo complexity of purification with the complexities of purification of the individual reduced density matrices of the states $|\psi_1\rangle$ and $|\psi_2\rangle$. The main findings of this work are listed below. 

\begin{enumerate}
           \item Although the pseudo complexity is supposed to be a simple sum of the two states involved in the construction of the transition matrix, the pseudo complexity of purification can show non-trivial behaviour compared to the individual  complexities of purification associated with the reduced density matrices of the two states involved. 
           
           \item We also find that the mutual pseudo complexity of purification satisfies the generally expected inequality $\Delta C(\tau_A^{1|2})\leq 0$ \cite{Caceres:2019pgf} \footnote{Note that the equality sign is different from \cite{Caceres:2019pgf} due to our choice of definition.}. Also, for all scaling exponents and masses corresponding to the two states $|\psi_1\rangle$ and $|\psi_2\rangle$, the saturation of the pseudo  complexity of purification is similar to the saturation of half the sum of individual  complexities of purification for the reduced density matrices $\rho_A^1$ and $\rho_A^2$ associated to the initial and final states. More details on the behaviour of mutual pseudo complexity of purification are there in Sec.~(\ref{mutualcop}).
           
           \item From the behaviour of pseudo complexity of purification, it appears that the qualitative behaviour remains similar to that of the usual complexity of purification of a reduced density matrix. Hence, we expect the general dependence of the quantity in terms of the system parameters to be similar to that of the complexity of purification of the reduced density matrices. However, the pseudo complexity of purification, since it carries information of both the states $|\psi_1\rangle$ and $|\psi_2\rangle$, is expected to depend on parameters of two states instead of one. This also results in an increment in the number of auxiliary parameters to be added and minimized for finding the most optimal purification. As we find, the dependence on the two states should be symmetric. It is not surprising since the transition matrix, or more specifically, the reduced transition matrix by structure, does not differentiate in any way between the two states used to define it.
           
           \item For the most general purification of the reduced transition matrix with $n$ oscillators, the number of unknown parameters scales as $2n^2$. This may be understood from eqs. \eqref{JKL_eqs}; for a reduced system made of $n$ linear harmonic oscillators, the $J$, $K$, and $L$ matrices would be square matrices of rank $n$. The choice \eqref{JKL_eqs} always leave any two of the six matrices undetermined, and thus the number of unknown parameters in the final optimization problem is $2n^2$.
                      
\end{enumerate}

Our analysis is mostly done for a reduced transition matrix of one oscillator after integrating the rest. This analysis can be extended to more number of reduced oscillators in many ways. However, due to the increase of unknown parameters, the complexity optimization problem becomes numerically more challenging, in general. However, one can perform a mode-by-mode purification as done in \cite{Caceres:2019pgf} on a physical basis. There, instead of optimizing all the modes together, one purifies each of the modes individually and then optimizes the sum of the complexities for all the modes. This decreases the number of unknowns significantly. For mode-by-mode purifications, the number of unknowns will scale as just $2n$ as each mode involves $2$ auxiliary parameters. However, since we get similar quantitative behaviour for pseudo complexity of purification as usual complexity of purification for the reduced system of one oscillator, we expect this to hold for the reduced system with more number of modes. With this assumption, we can therefore use a similar form for the pseudo complexity of purification, in terms of variables of two states instead of one 
\begin{equation}
	\mathcal{C}_P\left(\tau_{A}^{1|2}\right) = \frac{N}{2} \log\left(\omega_2\,\delta \right) + \frac{1}{2} \log\left(\frac{1}{f_1\left(m_1, z_1, m_2, z_2\right) N \delta} \right) - \frac{f_2\left(m_1, z_1, m_2, z_2\right) N^2 \delta^2}{48} + \cdots, 
\end{equation}
where $\delta$ is the gap between two points on the lattice \footnote{This variable does not come in our study since we did not consider any disjoint systems.}, this expression should work for very small mass, and the ellipses denote higher order terms. Usually, $f_1 \sim \mathcal{O}\left(m_{1, 2}\right)$, and $f_2 \sim \mathcal{O}\left(m_{1, 2}^2\right)$. The functions $f_1$ and $f_2$ are symmetric under the exchange $1 \leftrightarrow 2$, but arbitrary otherwise \footnote{One possibility is $f_1 = \frac{1}{2}\sqrt{m_1^{2 z_1} + m_2^{2 z_2}}$ and $f_2 = f_1^2$.}. Apart from the symmetric dependence, the pseudo complexity of purification has to follow all the necessary properties as mentioned in Sec.~(\ref{sec3}) for both small and large mass regime. Also, the nontrivial dependence on the scaling exponents have to abide by the plots we have found and reported in the previous section. A more detailed study for subsystems with considering larger size (by taking mode by mode purification), disjoint nature etc. can yield more exact behaviour which is a very natural extension to this work. 

There are a bunch of interesting questions one can ask in this direction. Since the post-selected states might be, in general result of a non-unitary evolution, it might be related to studies of complexity for open systems where the Lindbladian evolution is non-unitary. One recent study in this direction is \cite{Bhattacharya:2022gbz}, where initial steps towards studying Krylov complexity have been taken. It would be interesting if one could relate the notion of Nielsen complexity with the Krylov complexity along these lines. Another natural direction is to study the pseudo complexity of purification for spin systems, e.g. in the transverse field Ising model and check if the complexity can probe phase transitions like pseudo entropy \footnote{This can be further compared with the study of entanglement of purification for transverse field Ising model as done in \cite{Bhattacharyya:2019tsi} as well as with the corresponding study of phase transitions using circuit complexity. Interested readers are referred to \cite{Ali:2018aon}  and the citations of it for the application of circuit complexity to detect phase transitions in various quantum systems.}. It would also be interesting to study conformal systems as done for usual  complexity of purification in \cite{Camargo:2020yfv}. Finally, some holographic notions of pseudo entropy by computing minimal surfaces in Euclidean setups were proposed in \cite{Nakata:2020luh}. Therefore, it would be natural to compute the volumes below those minimal surfaces and compute a holographic pseudo subregion complexity (similar to holographic subregion complexity \cite{Bhattacharya:2019zkb, Bhattacharya:2020uun, Bhattacharya:2021jrn, Bhattacharya:2021dnd, Bhattacharya:2021nqj}) in those geometries. We hope to address some of these problems later. 


\providecommand{\href}[2]{#2}\begingroup\raggedright\endgroup

\end{document}